\documentclass{article}

\usepackage{meta}
\usepackage{amssymb}
\usepackage{amsmath}
\usepackage{epsfig}
\usepackage{stfloats}

\title{FDTD subcell graphene model beyond the thin-film approximation}

\makeatletter
\def\name#1{\gdef\@name{#1\\}}
\makeatother
\name{{\bf \large  Ilya Valuev$^{1,2}$, Sergei Belousov$^{1,3*}$, Maria Bogdanova$^{1}$,}\\
      {\bf \large  Oleg Kotov$^4$, and Yurii Lozovik}$^4$}
\address{
$^1$Kintech Lab Ltd., 3rd Khoroshevskaya st. 12, Moscow, 123298, Russia \\
$^2$Joint Institute for High Temperatures of the Russian Academy of Sciences (JIHT), Izhorskaya st. 13, Bd.2, Moscow, 125412, Russia  \\
$^3$National Research Center `Kurchatov Institute', Akademika Kurchatova sq., 1,  Moscow, 123182, Russia \\
$^4$Institute of Spectroscopy of the Russian Academy of Sciences (ISAN), Fizicheskaya Str., 5, Troitsk, Moscow, 108840, Russia \\
*corresponding author, E-mail: {\tt belousov@kintechlab.com}
}

\begin{document}
\maketitle

\begin{abstract}
A subcell technique for calculation of optical properties of graphene with the finite-difference time-domain (FDTD) method is presented. The technique takes into account the surface conductivity of graphene which allows the correct calculation of its dispersive response for arbitrarily polarized incident waves interacting with the graphene.
The developed technique is verified for a planar graphene sheet configuration against the exact analytical solution. Based on the same test case scenario, we also show that the subcell technique demonstrates a superior accuracy and numerical efficiency with respect to the widely used thin-film FDTD approach for modeling graphene. We further apply our technique to the simulations of a graphene metamaterial containing periodically spaced graphene strips (graphene strip-grating) and demonstrate good agreement with the available 
theoretical results.

\end{abstract}

\section{Introduction}

Graphene has emerged as a candidate for a large number of novel applications in the field of nano-optics. Unique optical properties of graphene are determined by a 2D nature of its conductivity, with one of the most notable consequences being the existence of ultra-subwavelength graphene surface plasmon polaritons on its surface in the THz frequency range ~\cite{Geim}. Coupled with the fact that the conductivity of graphene is easily controlled by electrical gating, this effect can be utilized to create tunable graphene-based THz metamaterials~\cite{TunableTHZ}.

Though the electromagnetic problem for the continuous graphene sheet can be solved analytically~\cite{Falkovsky,Stauber}, in most cases graphene-based metamaterials with complicated geometry require numerical methods. Nowadays, there are three popular computational electromagnetic methods for graphene simulation: finite-difference time-domain method (FDTD), finite element method (FEM), and method of moments (MoM). Each method has its own advantages and disadvantages depending on the specific problem (see the review~\cite{RevMethods}). Compared to frequency-domain methods FDTD being time-domain has several advantages, among which less computing memory and simulation time for obtaining spectra in a broad frequency range. There are different approaches to model graphene in FDTD. In the most common approach, 2D graphene sheet is represented on a 3D mesh by a thin film with its thickness of the order of a single mesh step. For the wavelengths of interest in the THz range (100 microns) the mesh step should be of the order of several microns. However, it was shown that in order to provide a satisfactory approximation of a 2D sheet conducting properties, the mesh step should be superfine (tens of nanometers or less)~\cite{thinfilmappr,SPPerror,FresnelPRA16}, which makes the task computationally challenging. 
To overcome this problem some modified FDTD methods for graphene have been developed, among which the locally one-dimensional method~\cite{LOD-FDTD}, 
various subcell methods~\cite{subcell,subcellPML}, and the surface boundary condition method~\cite{boundary-FDTD}.

In this paper we present a subcell technique for conducting 2D materials which 
employs a tensor representation of the frequency-dependent dielectric permittivity. 
This technique was developed for interfaces of dispersive materials~\cite{smoothing} and implemented in the framework of EMTL library~\cite{EMTL}.
Unlike other subcell methods it formally allows for arbitrary orientation of the interface normal, not requiring the interface to
be aligned with the FDTD grid.
We verify the proposed technique by calculating the optical properties of a free-standing continuous graphene sheet at oblique incidence against the analytical solution. We then perform a head-to-head comparison of the subcell technique with the thin-film approach in identical simulation setups. We show that the proposed technique accurately accounts for the interaction of arbitrarily polarized light with the graphene sheet, in contrast to the thin-film approach.
We further compute the transmittance, reflectance and absorbance of a graphene strip-grating metamaterial at oblique incidence and demonstrate an excellent agreement of the results with the available semi-analytic solution~\cite{StripGrating}.

\section{Analytical solution for a conducting graphene sheet}

\begin{figure}[t!]
\centerline{
\epsfig{figure=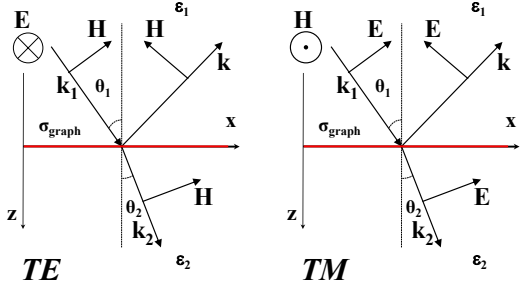,width=75mm}
}
\caption{Oblique incidence on a graphene sheet (highlighted in red). TE- and TM-polarization cases.}
\label{fig:tmte}
\end{figure}

Surface conductivity of graphene in IR and THz frequency ranges can be described with a single Drude-like term:

\begin{equation}
\sigma_{graph} = \frac{i e^2 E_F}{\pi\hbar^2 c}\frac{1}{\omega+i\Gamma},
\label{scond}
\end{equation}
where $e$ is the electron charge, $c$ is the speed of light in vacuum, $E_F$ is the Fermi energy of electrons in graphene, $\omega$ is the frequency, and $\Gamma$ is the phenomenological electron relaxation rate (damping)~\cite{GrapheneCond}. $E_F$ can be controlled by applying the gate voltage to the graphene, which in turn leads to the change of the conductivity $\sigma_{graph}$~\cite{TunableTHZ}.

\begin{figure*}[t!]
\centerline{
\epsfig{figure=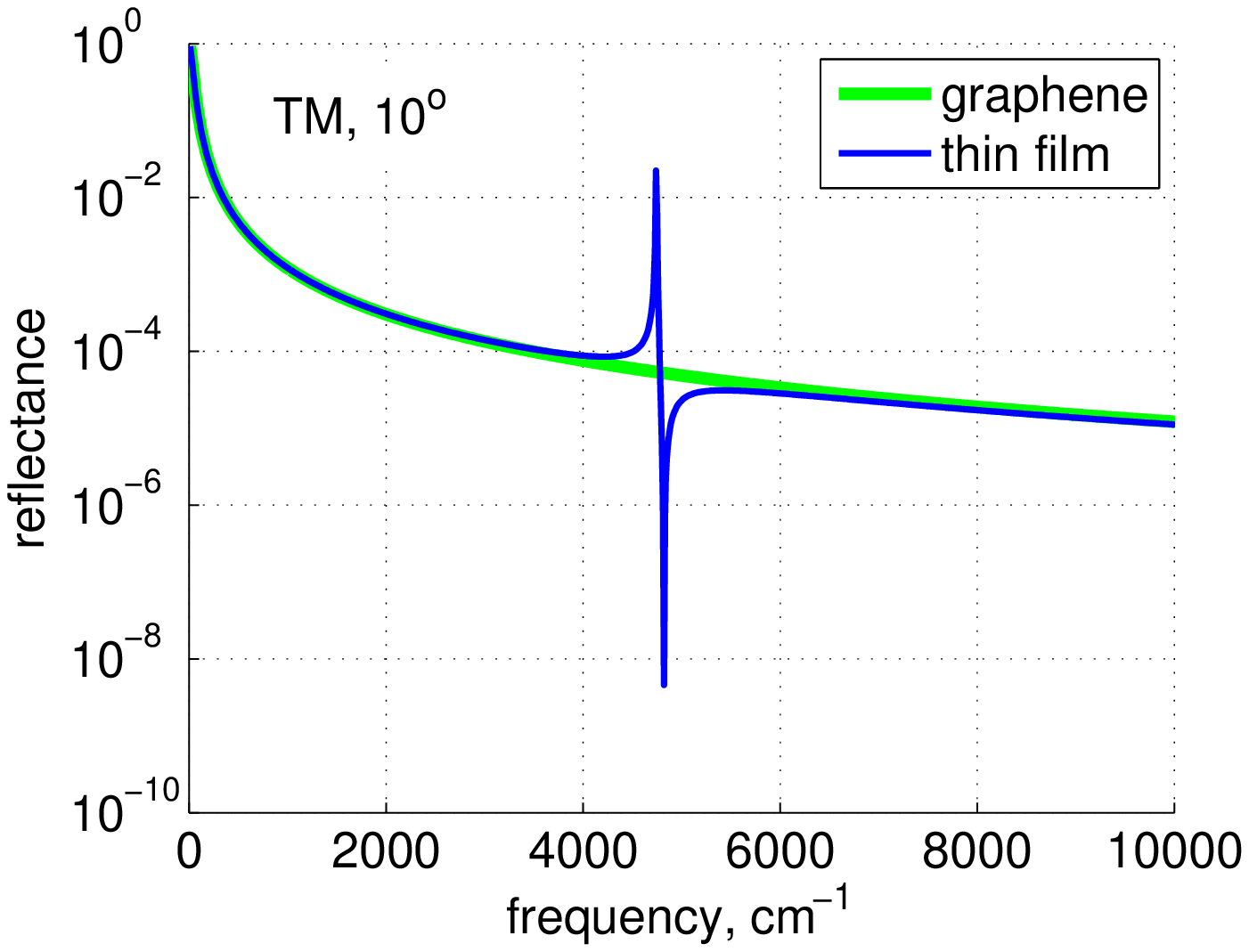,width=53mm}
\epsfig{figure=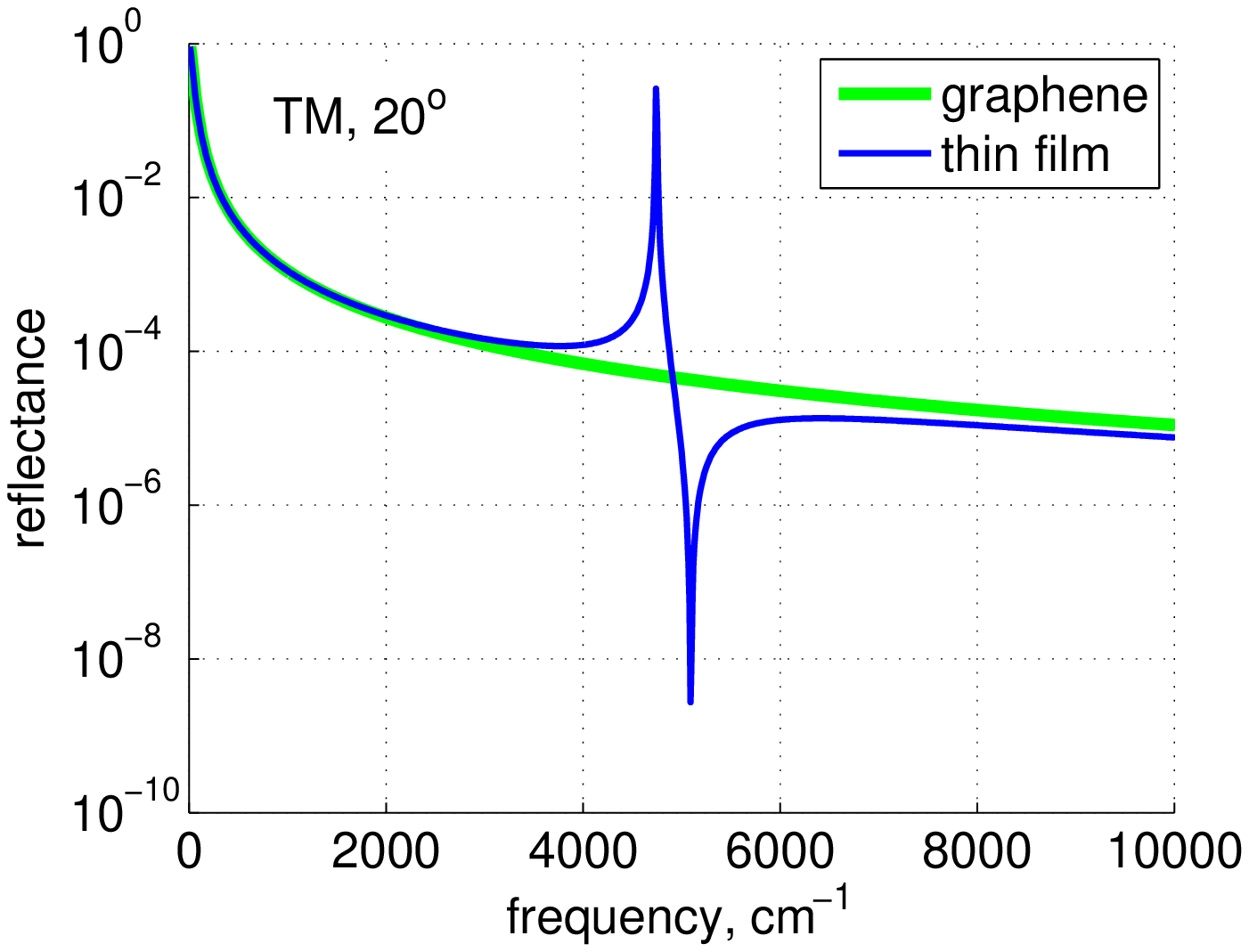,width=53mm}
\epsfig{figure=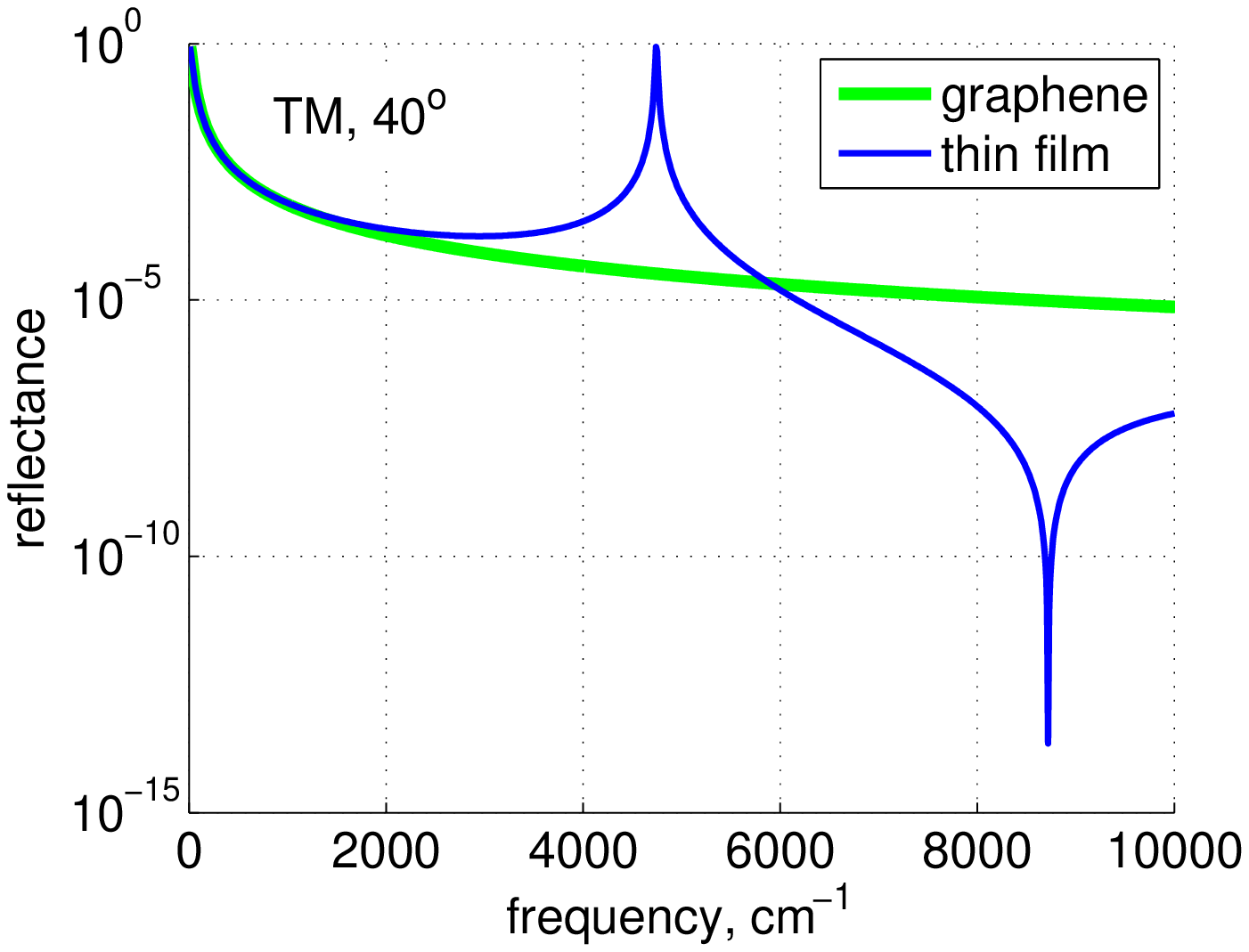,width=53mm}
}
\centerline{
\epsfig{figure=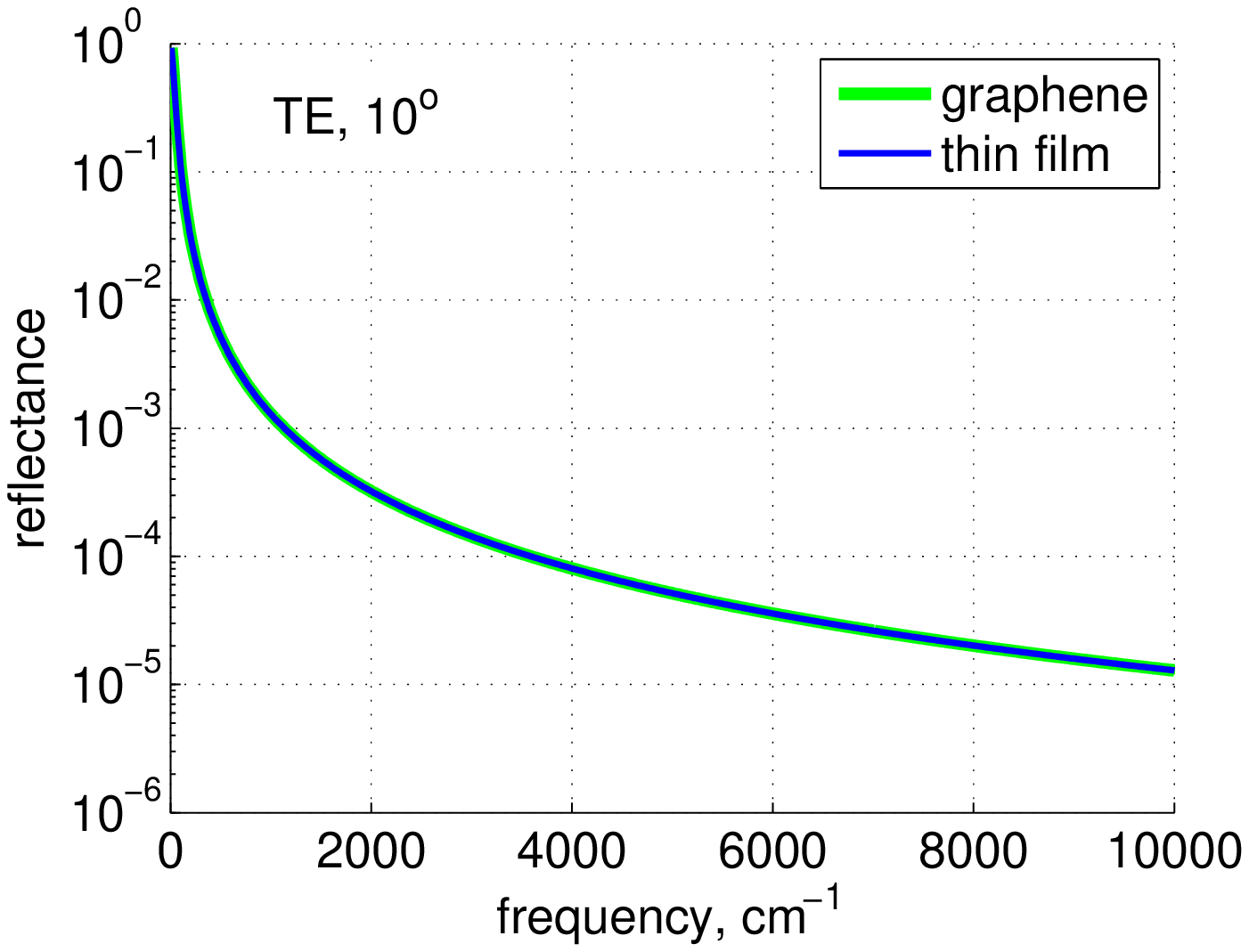,width=53mm}
\epsfig{figure=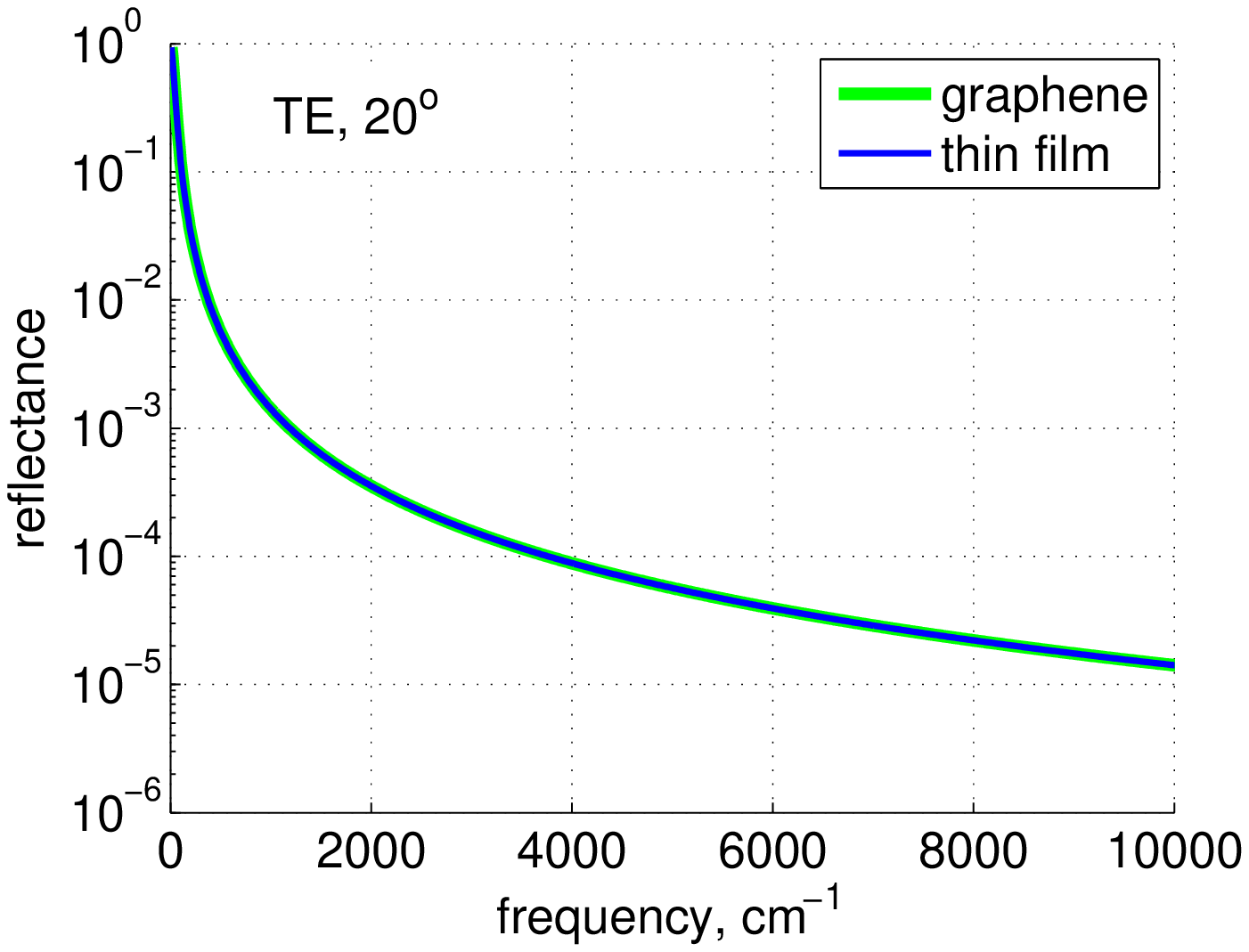,width=53mm}
\epsfig{figure=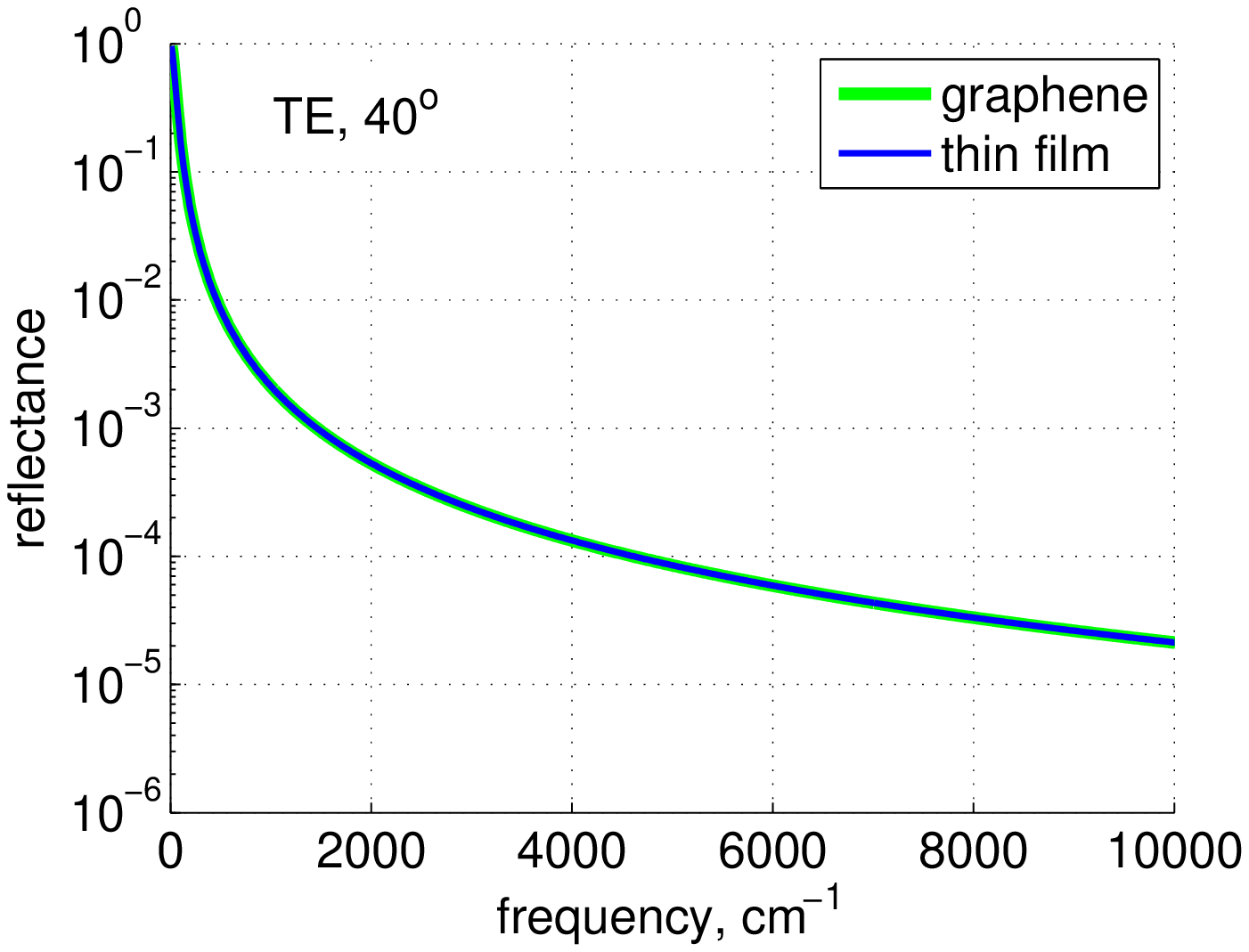,width=53mm}
}
\caption{Reflectance for the free-standing graphene sheet ($E_F = 0.3$ eV, $\Gamma = 0$) calculated using the exact analytical solution and compared to the thin-film approach with the effective dielectric function~(\ref{eps}).}
\label{fig:plate}
\end{figure*} 

Let us consider an oblique incidence of a plane wave on a continuous graphene sheet. The graphene sheet sandwiched between two half-spaces $1$ and $2$ filled with dielectric media $\varepsilon_1$ and $\varepsilon_2$ respectively, as shown in~Fig.~\ref{fig:tmte}. The respective wavevectors of the wave in the half-spaces $1$ and $2$ are ${\bf k}_{1,2} = \frac{\omega}{c}\sqrt{\varepsilon_{1,2}}$. The incident wave gives rise to surface currents at the sheet, which can be expressed as: 

\begin{equation}
{\bf J}_{graph} = \sigma_{graph}{\bf E}(0),
\label{scur}
\end{equation}
where ${\bf J}_{graph}$ is the surface current density at the graphene sheet, and ${\bf E}(0)$ is the electric field at the graphene sheet. Taking into account the relation~(\ref{scur}), we can write the boundary conditions for the electric (${\bf E}$) and magnetic (${\bf H}$) fields in case of TE- and TM-polarized incident waves respectively:

\begin{equation}
\begin{aligned}
E_{1y} = E_{2y} = E_y(0),\\
\mu_1 H_{1z} = \mu_2 H_{2z},\\
H_{2x} - H_{1x} = \frac{4\pi}{c} J_y = \frac{4\pi \sigma_{graph}}{c} E_y(0),
\end{aligned}
\label{BndTE}
\end{equation}


\begin{equation}
\begin{aligned}
E_{1x} = E_{2x} = E_x(0),\\
({\bf D}_2 - {\bf D}_1)\cdot {\bf n} = 0, \\ 
H_{2y} - H_{1y} = \frac{4\pi}{c} J_x = \frac{4\pi \sigma_{graph}}{c} E_x(0),
\end{aligned}
\label{BndTM}
\end{equation}

These boundary conditions are then used to derive the Fresnel coefficients $r_g$ and $t_g$ of a graphene sheet sandwiched between the two dielectric half-spaces~\cite{Falkovsky,Stauber}. In case of the TE-polarized incident wave we obtain:

\begin{equation}
\begin{aligned}
r_g^{TE} = \frac{{\left( {\sqrt {\frac{{{\varepsilon _1}}}{{{\mu _1}}}} \cos {\theta _1} - \sqrt {\frac{{{\varepsilon _2}}}{{{\mu _2}}}} \cos {\theta _2} - \frac{{4\pi \sigma }}{c}} \right)}}{{\left( {\sqrt {\frac{{{\varepsilon _1}}}{{{\mu _1}}}} \cos {\theta _1} + \sqrt {\frac{{{\varepsilon _2}}}{{{\mu _2}}}} \cos {\theta _2} + \frac{{4\pi \sigma }}{c}} \right)}},\\
t_g^{TE} = \frac{{2\sqrt {\frac{{{\varepsilon _1}}}{{{\mu _1}}}} \cos {\theta _1}}}{{\left( {\sqrt {\frac{{{\varepsilon _1}}}{{{\mu _1}}}} \cos {\theta _1} + \sqrt {\frac{{{\varepsilon _2}}}{{{\mu _2}}}} \cos {\theta _2} + \frac{{4\pi \sigma }}{c}} \right)}},
\end{aligned}
\label{FreTE}
\end{equation}
where $\theta_1$ is an incident angle, and $\theta _2$ is a transmittance angle.

For TM-polarized wave these coefficients take form:
\begin{equation}
\begin{aligned}
r_g^{TM} = \frac{{\left( {\sqrt {\frac{{{\varepsilon _2}}}{{{\mu _2}}}} \cos {\theta _1} - \sqrt {\frac{{{\varepsilon _1}}}{{{\mu _1}}}} \cos {\theta _2} + \frac{{4\pi \sigma }}{c}\cos {\theta _1}\cos {\theta _2}} \right)}}{{\left( {\sqrt {\frac{{{\varepsilon _2}}}{{{\mu _2}}}} \cos {\theta _1} + \sqrt {\frac{{{\varepsilon _1}}}{{{\mu _1}}}} \cos {\theta _2} + \frac{{4\pi \sigma }}{c}\cos {\theta _1}\cos {\theta _2}} \right)}}, \\
t_g^{TM} = \frac{{2\sqrt {\frac{{{\varepsilon _1}}}{{{\mu _1}}}} \cos {\theta _1}}}{{\left( {\sqrt {\frac{{{\varepsilon _2}}}{{{\mu _2}}}} \cos {\theta _1} + \sqrt {\frac{{{\varepsilon _1}}}{{{\mu _1}}}} \cos {\theta _2} + \frac{{4\pi \sigma }}{c}\cos {\theta _1}\cos {\theta _2}} \right)}}
\end{aligned}
\label{FreTM}
\end{equation}

The formulae~(\ref{FreTE}) and~(\ref{FreTM}) can be used to obtain analytically the reflectance and transmittance properties for the considered geometry for any given angle of incidence and the conductivity of the graphene.

An alternative so called 'thin-film' approach to the the problem depicted in~Fig.~\ref{fig:tmte} consists in substituting the graphene surface with a thin film. In this case, the thin film possesses an effective dielectric function, which mimics the properties of the graphene~\cite{thinfilmappr}:

\begin{equation}
\varepsilon (\omega)  = 1 + \frac{{4\pi i}}{\omega }\frac{{{\sigma _{graph}}(\omega)}}{{{d_{film}}}}
\label{eps}
\end{equation}
where $d_{film}$ is a film thickness. The reflectance and transmittance can then be calculated analytically with Airy formula.

In order to estimate the range of applicability of the thin film approach we now consider a special case of a free-standing continuous graphene sheet possessing a typical value of $E_F = 0.3$ eV. In our case the electron relaxation rate $\Gamma\ll E_F$ is of the order of several meV~\cite{GrapheneCond} and it slightly affects the results, so we can set $\Gamma = 0$.
Let us compare the exact analytical solution~(\ref{FreTE}) and~(\ref{FreTM}) for this case with the thin-film approach. The results are presented in~Fig.~\ref{fig:plate} for a range of the incident angles. In our calculations we took a typical thickness $d_{film} = 5$ nm, which is used in FDTD simulations of the graphene~\cite{thinfilmappr}.

We immediately see that while the thin-film approach gives accurate results in case of TE-polarization, it yields incorrect results in case of TM-polarization. Particularly, an unphysical resonance appears which becomes stronger as the incident angle is increased. The reason for this resonance to occur lies in unphysical normal current components arising within the thin film when the electric field in the incident wave has non-zero normal components (which is the case in TM-polarized wave). In the subsequent sections we describe the subcell technique which allows to avoid this effect.



\section{Subcell graphene model}

As shown in the previous section, the thin-film representation of graphene leads
to substantial differences in the reflection and transmission of the sheet compared to the actual 2D material.
In~\cite{smoothing} we used the surface normal direction to construct an effective
dielectric tensor for mesh cells in the vicinity of an interface. In this approach,
the graphene sheet may be regarded as such an interface, introducing additional dispersion
along the surface directions. Following the notation from~\cite{smoothing}, the effective inverse dielectric permittivity
tensor has the following form~\cite{VP-EP,tensor,tensor2,Johnson}:

\begin{equation}
\hat{\bf \epsilon}^{-1}={\bf P}<\epsilon^{-1}>+({\bf 1}-{\bf
P})<\epsilon>^{-1},
\label{tensor}
\end{equation}
where ${\bf P}$ is the projection matrix $P_{ij}=n_in_j$ onto the
normal $\vec{n}$ to the graphene sheet. Here the averaging $<>$ should be performed
 over the characteristic mesh cell volume $\delta v$, and assuming uniform bulk medium
it may be obtained analytically:

\begin{equation}
\begin{aligned}
\hat{\bf \epsilon}^{-1}=\\
{\bf P}\epsilon_{\rm bulk}^{-1}+({\bf 1}-{\bf
P})[({\delta v_{\rm graph}})^{\frac{1}{3}}\epsilon_{||}^{-1}+(1-\frac{\delta v_{\rm graph}}{\delta v})\epsilon_{\rm bulk}^{-1}],
\end{aligned}
\label{tensor1}
\end{equation}
where $\epsilon_{||}$ is the 2D graphene permittivity, and ${\delta v_{\rm graph}}$ is the volume fraction of the mesh cell,
"occupied" by the graphene sheet. The value of  ${\delta v_{\rm graph}}$ depends on the discretization thickness of the graphene sheet and the surface fraction of the sheet residing inside the cell
$\delta v_{\rm graph}=d_{\rm graph} s_{cell}$. Note that the discretization thickness $d_{\rm graph}$ may be chosen
smaller than the typical mesh step, thus reducing ${\delta v_{\rm graph}}$.  The solution of discrete
Maxwell's equations with the tensor~(\ref{tensor1}) is then obtained by the auxiliary
differential equation (ADE) method in the way described in~\cite{smoothing}.

Applying the subcell approach~(\ref{tensor1}) to the graphene sheets aligned with the (rectangular) Yee mesh is straightforward and leads to accurate solutions, as will be shown by the results presented in the next Section. In this case we select
$\delta v_{\rm graph}=\delta v$ and the electric field components aligned with the graphene sheet are updated with the
graphene dispersive ADEs, whereas the normal components are non-dispersive. Note that in the thin-film approach the
normal components of the electric field inside the sheet (represented by one or more mesh cells) are also governed by the dispersive equations, which appears to lead to inaccuracy in the solution.

Here we must remark that although the tensor method formally allows for arbitrary direction of the 2D interface,
its application to the inclined 2D sheets requires special care. Our preliminary tests
have shown, that variations in ${\delta v_{\rm graph}}$
between adjacent mesh cells, typical for the inclined geometry, lead to large discontinuities in the polarization currents
drastically reducing the accuracy. Therefore we leave the study of the inclined subcell model for future publications,
concentrating currently on the comparison of the thin film and the subcell approaches in the mesh-aligned case, but for the oblique wave incidence.

\section{Results and discussion}
\subsection{Continuous graphene} \label{Grsheet}

\begin{figure*}[t!]
\centerline{
\epsfig{figure=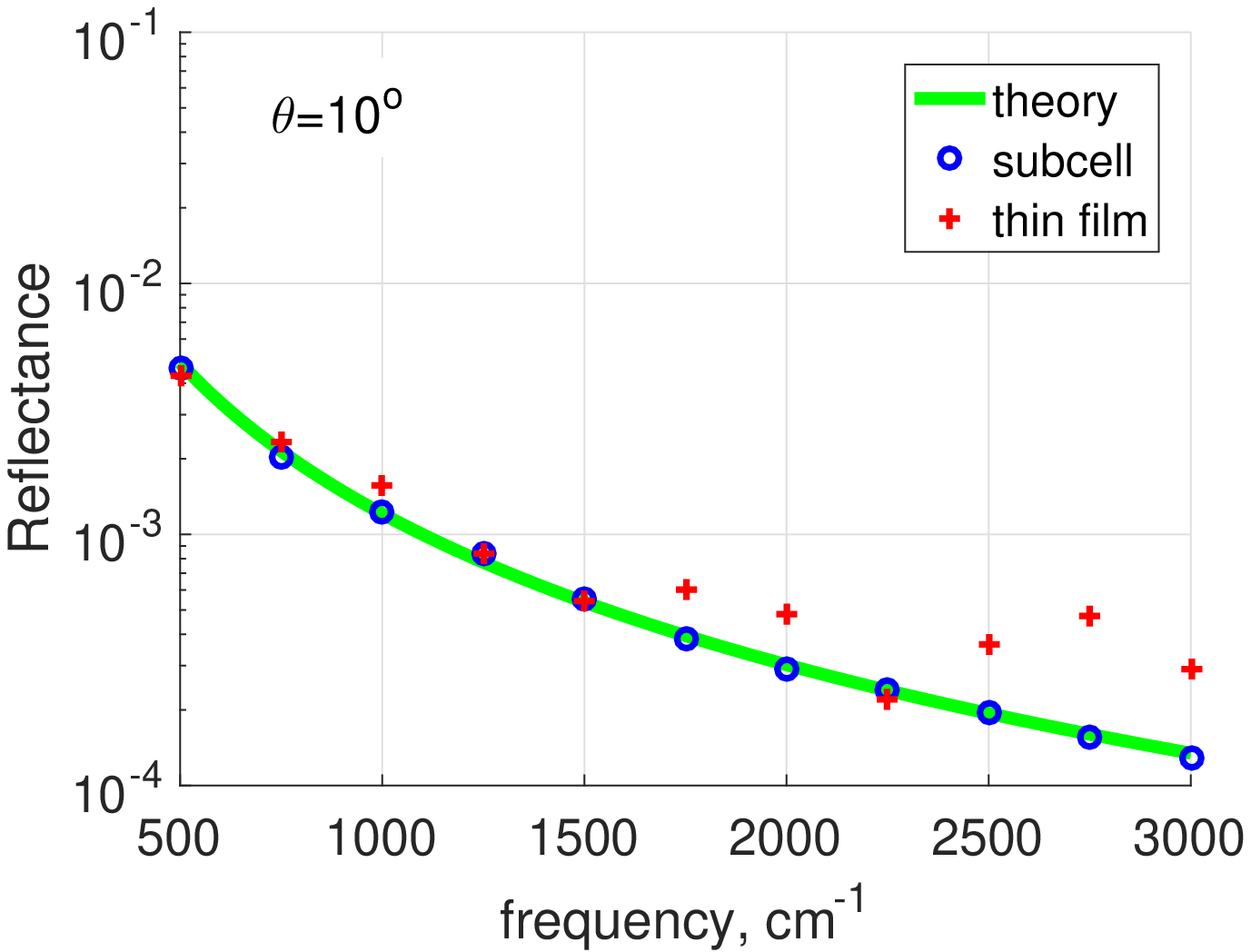,width=80mm}
\epsfig{figure=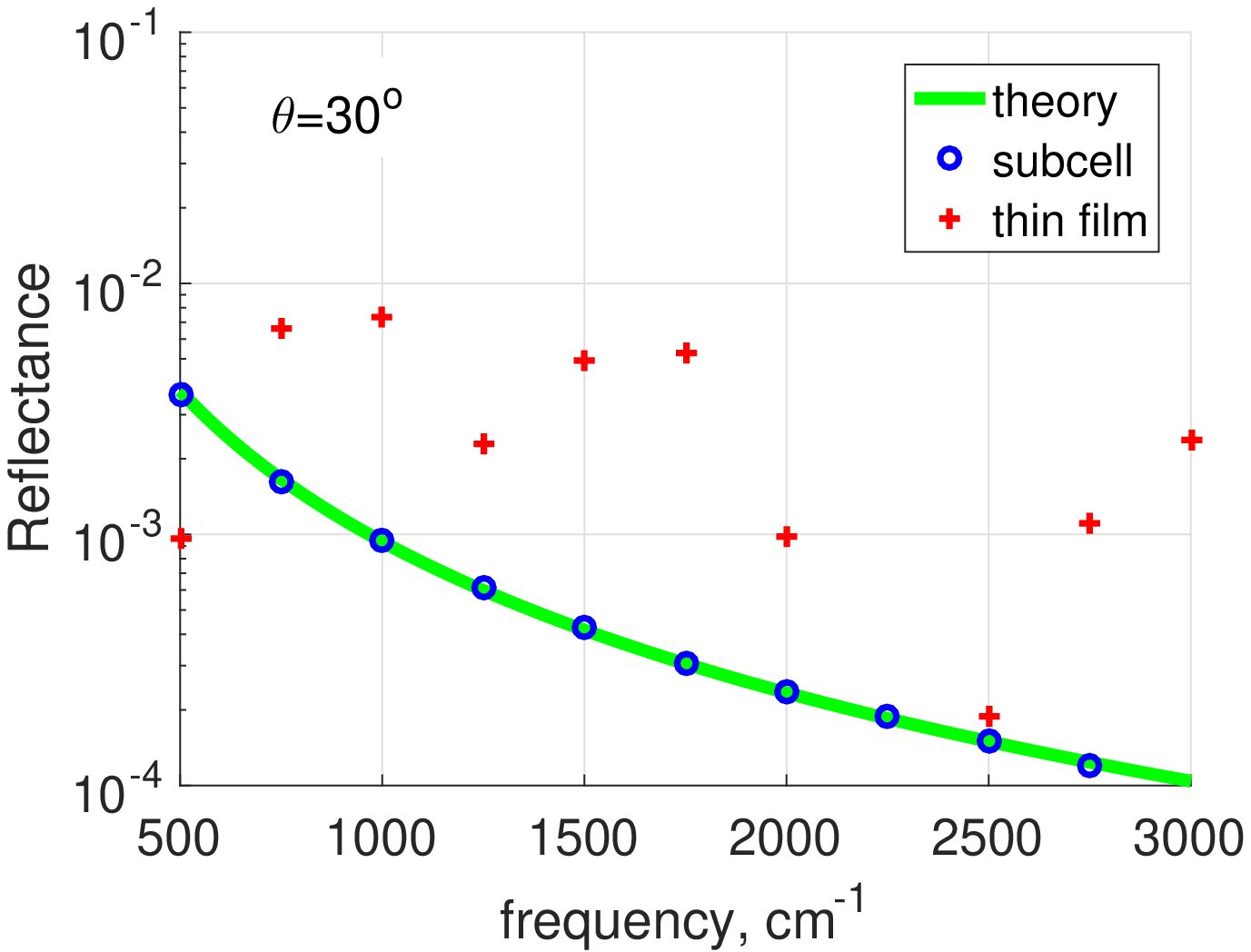,width=80mm}
}
\centerline{
\epsfig{figure=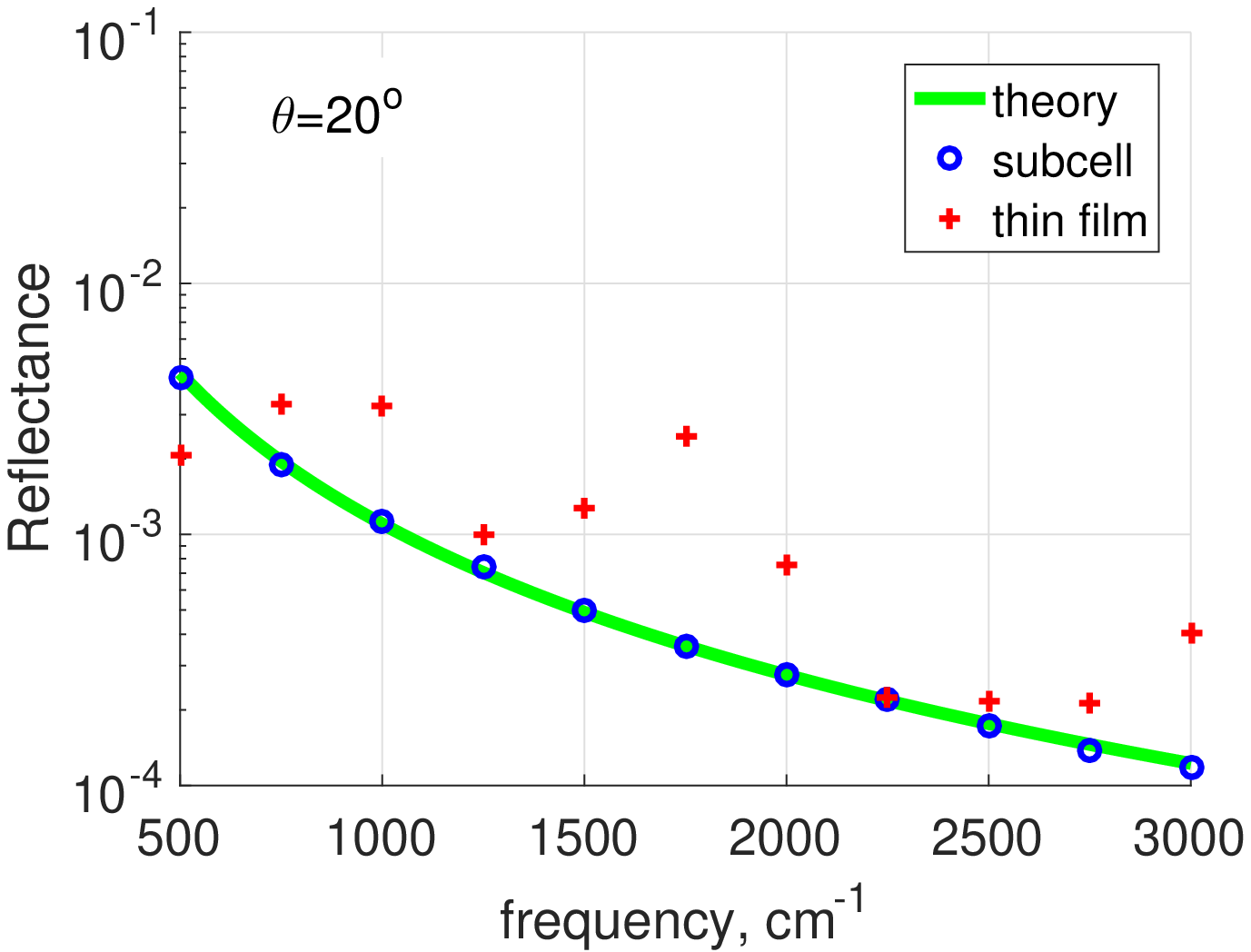,width=80mm}
\epsfig{figure=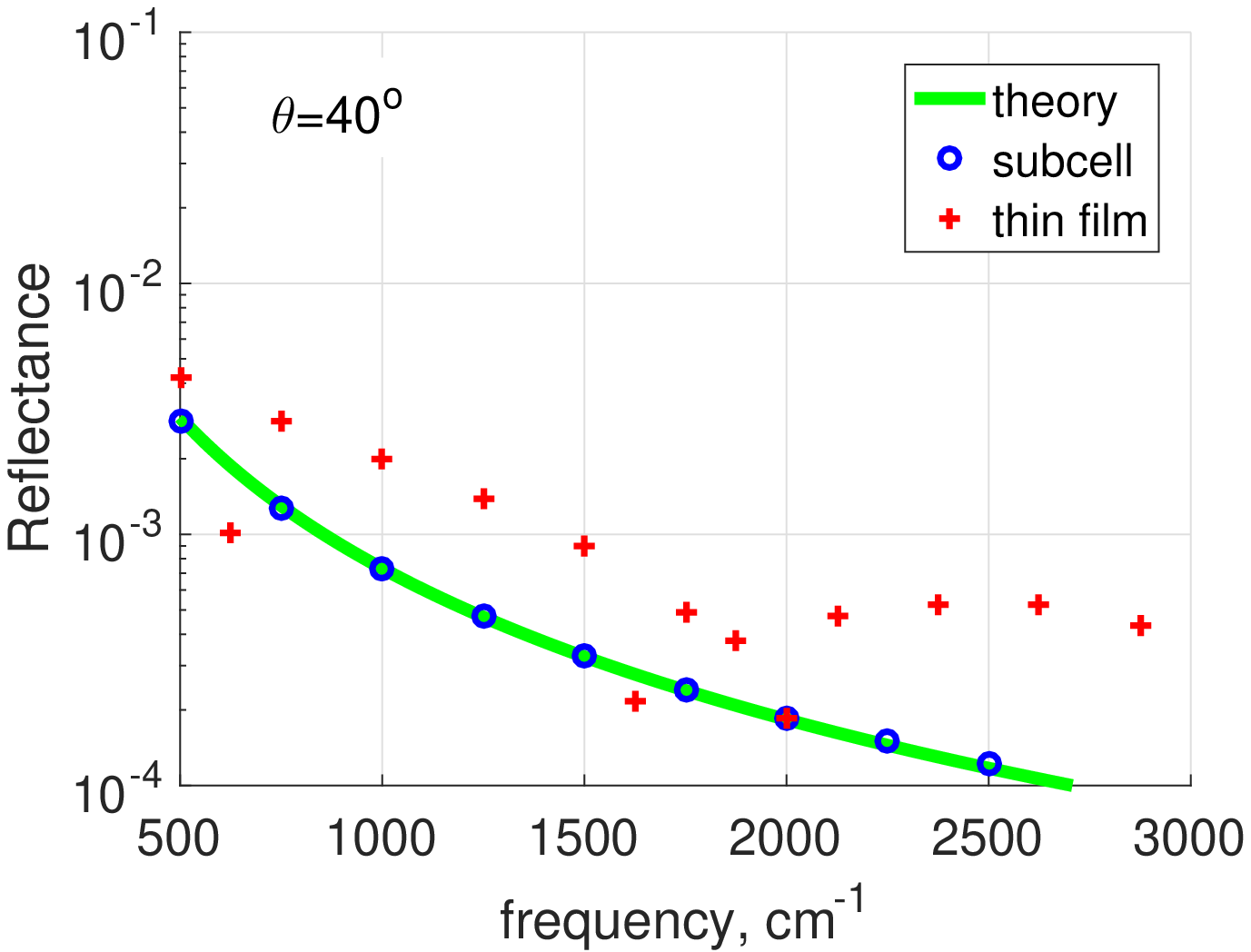,width=80mm}
}
\caption{FDTD calculation of reflectance for the continuous graphene sheet (TM-polarization), obtained using the subcell technique ($\Delta_{subcell}=50$ nm) and the thin-film approach ($d_{film}=5$ nm, $\Delta_{thin-film}=5$ nm).  Comparison is made with the analytical solution~((\ref{FreTE}) and~(\ref{FreTM})).}
\label{fig:fdtd}
\end{figure*}

In this section the FDTD simulation of oblique incidence of a plane wave on a graphene sheet in vacuum is described. The material parameters of graphene are the same as in the previous sections ($E_F = 0.3$ eV, $\Gamma = 0$). The comparison of the subcell technique with the thin-film approach is carried out. The results are also compared to the analytical solution~((\ref{FreTE}) and~(\ref{FreTM})).

In the simulations, the incoming oblique broadband plane wave impulse spans the near-IR to THz frequency range (500 --- 3000 $\text{cm}^{-1}$). We use the previously developed iterative oblique incidence technique~\cite{oblique} to obtain a broadband response for a given angle of incidence in a single simulation run.

In simulations with the developed subcell technique, the choice of the FDTD mesh step value is governed by the frequency range of interest as well as the geometrical features of considered structures. In our particular case of a planar graphene sheet the mesh step $\Delta$ value is determined solely by the frequency range. We found that the reflectance at a given frequency converges to within $2\cdot 10^{-3}$ at $\Delta_{subcell} = 50$ nm with respect to the result for $\Delta_{subcell} = 10$ nm. The results for TM-polarization are shown in~Fig.~\ref{fig:fdtd} (circles). The relative error of the reflectance calculations with respect to the analytical result is within $4\cdot 10^{-2}$ for the considered frequencies (below 3000 $\text{cm}^{-1}$), which corresponds to the absolute error less than $10^{-5}$ in the calculation of reflectance. Stability is ensured by performing prolonged simulations for $10^5$ time steps (1.7 ps). Typical simulations were run for $10^4$ time steps, which took about 10 s per one iteration of the oblique incidence method~\cite{oblique} at Intel Core i5 6200 processor.

\begin{figure}[t!]
\epsfig{figure=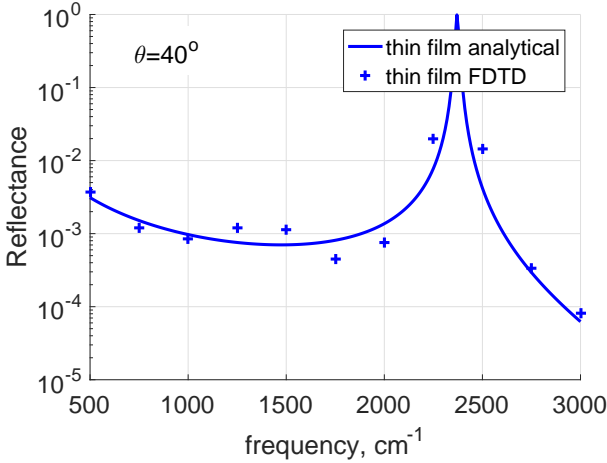,width=80mm}
\caption{FDTD calculation of reflectance for the continuous graphene sheet (TM-polarization), obtained using the thin-film approach. $d_{film}=20$ nm, $\Delta_{thin-film}=5$ nm. Comparison is made with the analytical solution (Airy formula) for the thin film.}
\label{fig:thinfilm20}
\end{figure}

On the other hand, the thin-film approach dictates the choice of a much more refined mesh step along the dimension normal to the graphene surface at least. In the reference simulations with the thin-film approach we used $\Delta_{thin-film} = 5$ nm, with $d_{film} = \Delta_{thin-film}$. This leads to a dramatic slow-down of the simulations by the factor of 240 with respect to the subcell technique (about 2400 s per one oblique iteration at Intel Core i5 6200 processor). The results and comparison with the subcell technique as well as the analytical solution are shown in~Fig.~\ref{fig:fdtd} (crosses). The relative error of the simulations with respect to the analytical result is greater than 0.1 for $\theta = 10$ degrees, rising to the order of 1 and even more for greater angles~(Fig.~\ref{fig:fdtd}).

The magnitude of error in case of the thin-film approach is due to two factors: a) the unphysical normal component of polarization current as discussed in previous section, b) the used mesh step is insufficient for the considered film thickness (single mesh step per film). In order to analyze the second factor, we computed the reflectance spectra of the graphene sheet assuming a thicker ($d_{film}=20$ nm) film in the thin-film approach, but leaving $\Delta_{thin-film} = 5$ nm. At $d_{film}/\Delta_{thin-film}=4$ mesh steps per film thickness, results appear to reproduce the analytical result for the thin film obtained with Airy formula ~(Fig.~\ref{fig:thinfilm20}). This result also indicates that a finer mesh is required in order to accurately reproduce the analytical thin-film result.


For TE polarization, the accuracy of the both methods is comparable. The subcell method is still much more preferable due to much higher computational performance.

\subsection{Graphene strip-grating}
\begin{figure}[t!]
\epsfig{figure=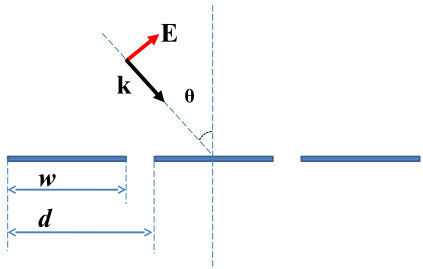,width=80mm}
\caption{Schematic of a graphene strip-grating metamaterial. Strips have the width $w$ and a 1D-periodicity with the period $d$. An oblique incidence of a TM-polarized wave with $\theta=30$ degrees is considered.}
\label{fig:stgratgeom}
\end{figure}

We further demonstrate the application of the subcell method to the modeling of a oblique incidence on a graphene metamaterial. We consider a graphene strip-grating~(Fig.~\ref{fig:stgratgeom}) possessing multiple surface plasmon resonances in THz range. These resonances can only be excited with a TM-polarized light (when the electric field has a component perpendicular to the strip axis)~\cite{TunableTHZ}. Dependence of the resonance positions on the width of the strips makes this structure suitable to serve as a tunable polarization-selective filter. The grating has the period of 70 $\mu$m and the strip width of 20, 40, and 60 $\mu$m. The ambient medium is vacuum. The TM-polarized plane wave is incident on the structure at an angle of 30 degrees.

While the width of the strips is of order of the wavelength of the impinging light, the arising resonances are strongly localized near the grating surface. Moreover, they possess very sharp field gradients at the edges of the strips~\cite{StripGrating}. This places strong requirements on the FDTD mesh step size, which should be as fine as 100 mesh steps per strip width at least~\cite{BelousovOLED}. The convergence tests for 60 $\mu$m strips at the normal incidence indicate that the transmittance spectra convergence to within 1\% relative error in 0---10 THz range at $\Delta = 250$ nm. This mesh step was used in the oblique incidence calculations.

\begin{figure*}[t!]
\centerline{
\epsfig{figure=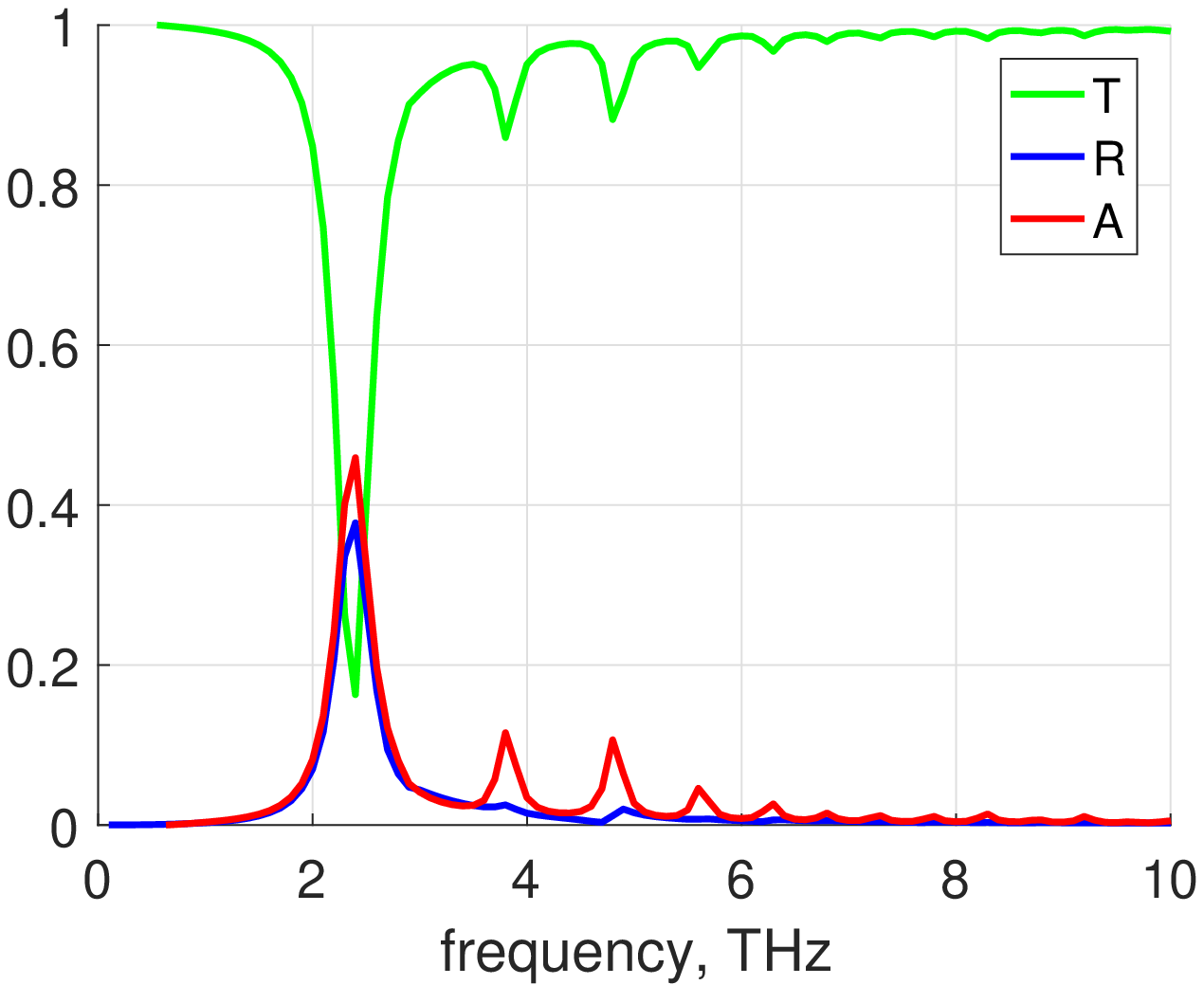,width=53mm}
\epsfig{figure=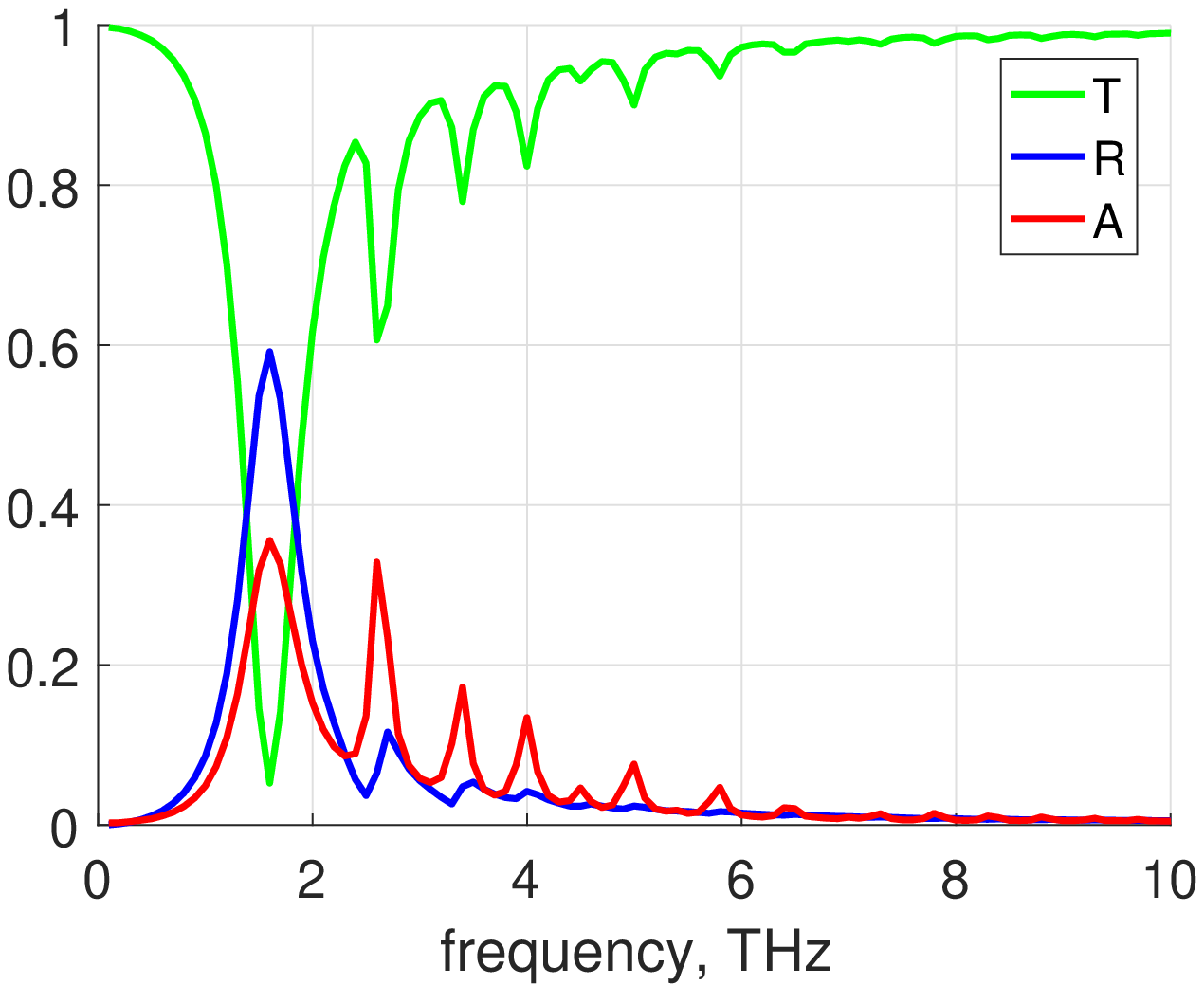,width=53mm}
\epsfig{figure=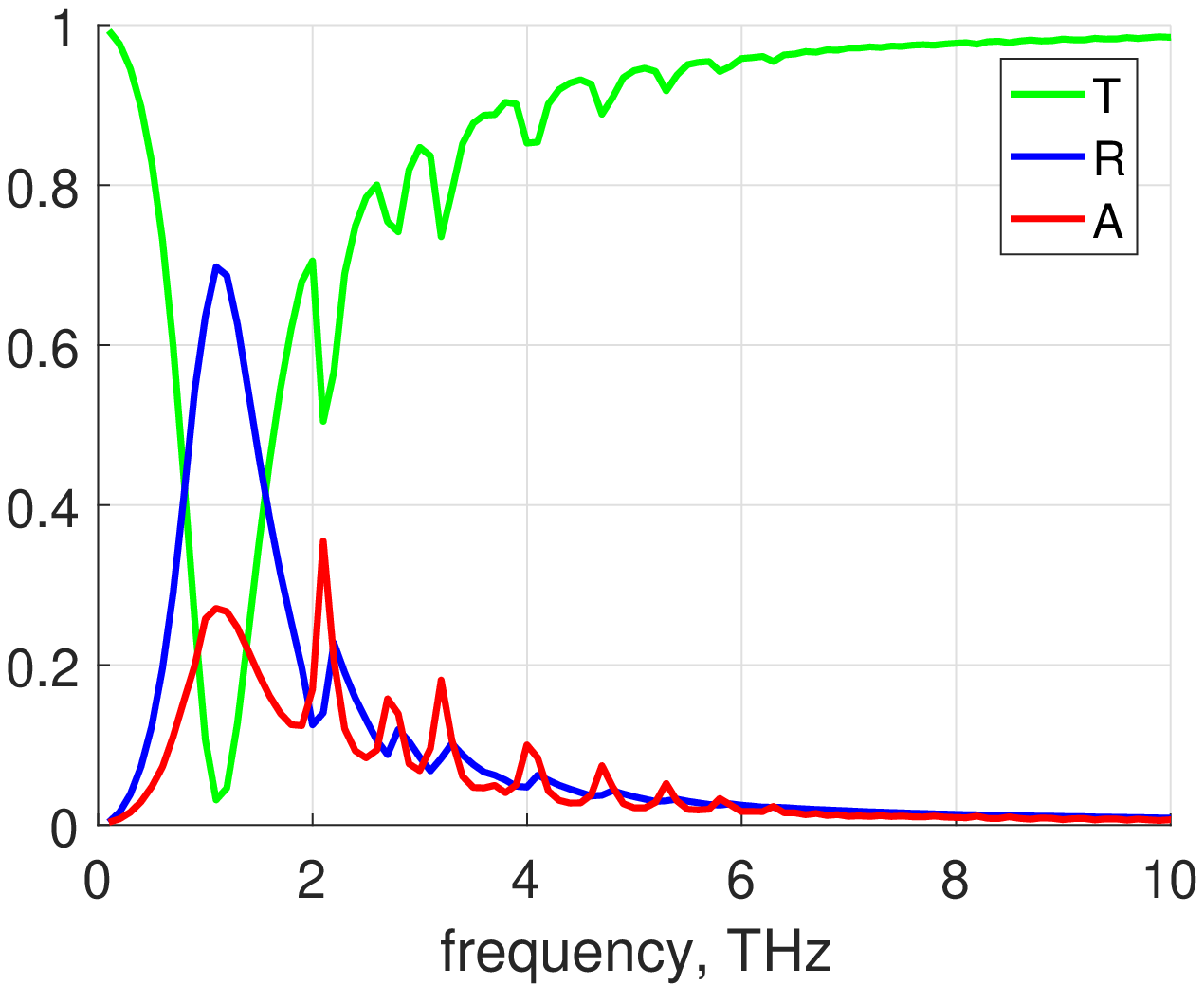,width=53mm}
}
\caption{FDTD calculation of transmittance (T), reflectance (R), and absorbance (A) for the graphene strip-grating, obtained using the subcell technique. The oblique incidence (30 deg) of a TM-polarized plane wave is considered. The parameters of the grating: period 70 $\mu$m, strip width 20 (left), 40 (center), and 60 $\mu$m (right). The graphene parameters: $E_F=0.39$ eV, $\gamma=0.66$ meV.}
\label{fig:strip}
\end{figure*}

The obtained spectra~(Fig.~\ref{fig:strip}) reveal the transmittance dips and absorbance peaks associated with the surface plasmon resonances. The results are consistent with the exact semi-analytical solution for this structure~\cite{StripGrating} based on the Fourier expansion method. However, this semi-analytical method assumes the numerical solution of the integral equations and requires using regularization techniques~\cite{StripGrating,RegIntEq}. Although such semi-analitical numerical methods allow modeling of more complicated structures than the strip-grating, such as corrugated graphene~\cite{CorrGraphene} or even biperiodic graphene metasurfaces~\cite{BiPeriodic}, for each new configuration they demand to rewrite all the equations and to rebuild the numerical solution. In contrast, the subcell method allows using FDTD as a universal package for the analysis of arbitrary geometrical configurations and propagation directions, including analysis of propagation along graphene sheets.

\section{Conclusions}

This paper presents a subcell technique for modeling graphene within the framework of FDTD method.
The proposed subcell technique accurately reproduces the surface conductivity of graphene in frame of the FDTD calculations, which is proved by simulating an oblique incidence of a plane wave on a free-standing continuous graphene sheet. It is shown that the subcell technique demonstrates a superior accuracy with respect to the thin-film approach. In particular, analysis of TM-polarization case shows that thin-film approach becomes inaccurate due to the unphysical polarization current component normal to the graphene surface. It is also shown, that the subcell method does not require a fine mesh leading to a drastic performance improvement of FDTD simulations of graphene containing structures as compared to the thin-film approach. 
The subcell technique is further verified by simulating a graphene strip-grating metamaterial at oblique incidence. It is shown that the results are consistent with the available semi-analytical solution.
We conclude that the subcell technique presents a flexible and efficient approach to modeling of graphene metamaterials and can be applied to optimize their optical properties for various applications.


\bibliographystyle{IEEEtran}

\end{document}